\documentclass[aip,author-year]{revtex4-1}

\usepackage{rotating}

\usepackage{amsbsy}
\usepackage{amssymb}
\usepackage{amsmath}
\usepackage{graphicx}
\usepackage{color}

\newcommand{\beq}{\begin{equation}}
\newcommand{\eeq}{\end{equation}}

\begin{document}

\title{Learning from the Frequency Content of Continuous Gravitational Wave Signals \footnote{To appear as a chapter in ``Astrophysics In The XXI Century With Compact Stars'', World Scientific, eds. Cesar Zen Vasconcellos and Fridolin Weber}}

\author{D. I. Jones}


\affiliation{Mathematical Sciences, University of Southampton}

\begin{abstract}

Spinning neutron stars can emit long-lived gravitational waves.  There are several mechanisms that can produce such \emph{continuous wave} emission.  These mechanisms relate to the strains in the elastic crust, the star's magnetic field, superfluidity of the neutron fluid, and bulk oscillations of the entire star.  In this chapter we describe how the frequency content of the gravitational wave signal, and its relation to any electromagnetically observed spin frequency, can be used to constrain the mechanism producing the gravitational waves.  These ideas will be of use in the event of the first detections of such signals, and help convert a detection into useful physical insight.

\end{abstract} 
  
\maketitle

\tableofcontents

\section{Overview}

Spinning neutron stars can emit gravitational waves, as long as their mass distribution is deformed away from axisymmetry in some way \citep{MTW_73, ande_19}.   Such gravitational wave signals could potentially last for very long periods of time.  The spin-down timescales of known pulsars give us an indication of these timescales, ranging from thousands of year for young pulsars, through to billions of years for millisecond pulsars (MSPs) (see e.g.\ \citet{lgs_12}).  Such durations are far longer than the duration of any gravitational wave observation, typically of length weeks to years, hence the designation of \emph{continuous waves} (CWs).   The gravitational wave signal itself may however change on timescales much shorter than this, if, for instance, there is a sudden release of strain within the star. This interplay of long and (possibly) short timescales makes searching for CWs computationally and theoretically challenging.

There has nevertheless been a sustained and vigorous effort to detect such CWs.  So far, no detections have been made, but upper limits have been put on such emission from both knows sources (pulsars and low-mass X-ray binaries), and from currently unknown sources.  See the following for recent examples: \citet{LVK_SNRs, LVK_Sco-X1, LVK_MSPs, LVK_all-sky}.  

There are several mechanisms that could produce such  deformations.  The two main scenarios are non-axisymmetries in rigidly rotating stars, colloquially referred to as \emph{mountains}, and \emph{oscillations} of the star; see \citep{gg_18_review} for a review.  However, within these two cases are multiple possibilities, relating to free precession \citep{zs_79, ja_02}, multi-component stars \citep{LVK_08_beating_Crab}, and superfluidity \citep{jone_10}, that can determine the strength and frequency spectrum of the emission. 

In this article, we look forward to the first detections, and ask the question as to what one can learn following such detections.  The physics relating to mountains and fluid oscillations combines many interesting areas, including nuclear physics, fluid dynamics (including magnetohydrodynamics), elasticity, superfluidity, superconductivity, and, of course, general relativity; see \citet{gg_18_review}.  There is clearly a lot at stake, in terms of turning a detection into physical insight.  This is a complex problem, and various facets of it can be found scattered over the extensive literature on CW emission.  Here we narrow this question down a little, and reflect upon the fact that (in some cases) different emission mechanisms produce gravitational waves at different frequencies.  We therefore pose  and attempt to answer the following question:
\begin{quotation}
\emph{In the event of the detection of a continuous gravitational wave source, to what extend can we use the frequency content of the signal to constrain the emission mechanism?}
\end{quotation}

Note that there are other candidate mechanisms for producing long-lived CW signals that have nothing to do with neutron stars, but these are of more speculative nature.  These include emission from boson clouds outside of black holes \citep{dant_18},  emission from low mass primordial black holes \citep{metal_21}, and the direct interaction of dark photon dark matter with the gravitational wave detectors \citep{LVK_21_DPDM}.  We will not consider such scenarios here.

The structure of this chapter is as follows.  In Section \ref{sect:mechanisms} I give a brief review of the main gravitational wave mechanisms that I will consider, describing their frequency content, and how it relates to an electromagnetically observed spin frequency.  In Section \ref{sect:single} I consider the scenario where a gravitational wave detection has been made, consisting of a single frequency, and describe what can be inferred, depending upon whether a radio pulsar spin frequency is know or not, and how that frequency compares with the gravitational wave frequency.  In Section \ref{sect:multi} I pose a similar question, but for the case where the are multiple frequency components in the gravitational wave signal.   I briefly summarise in Section \ref{sect:summary}.

\begin{sidewaysfigure}
\begin{center}
\includegraphics[height=9in, angle=-90]{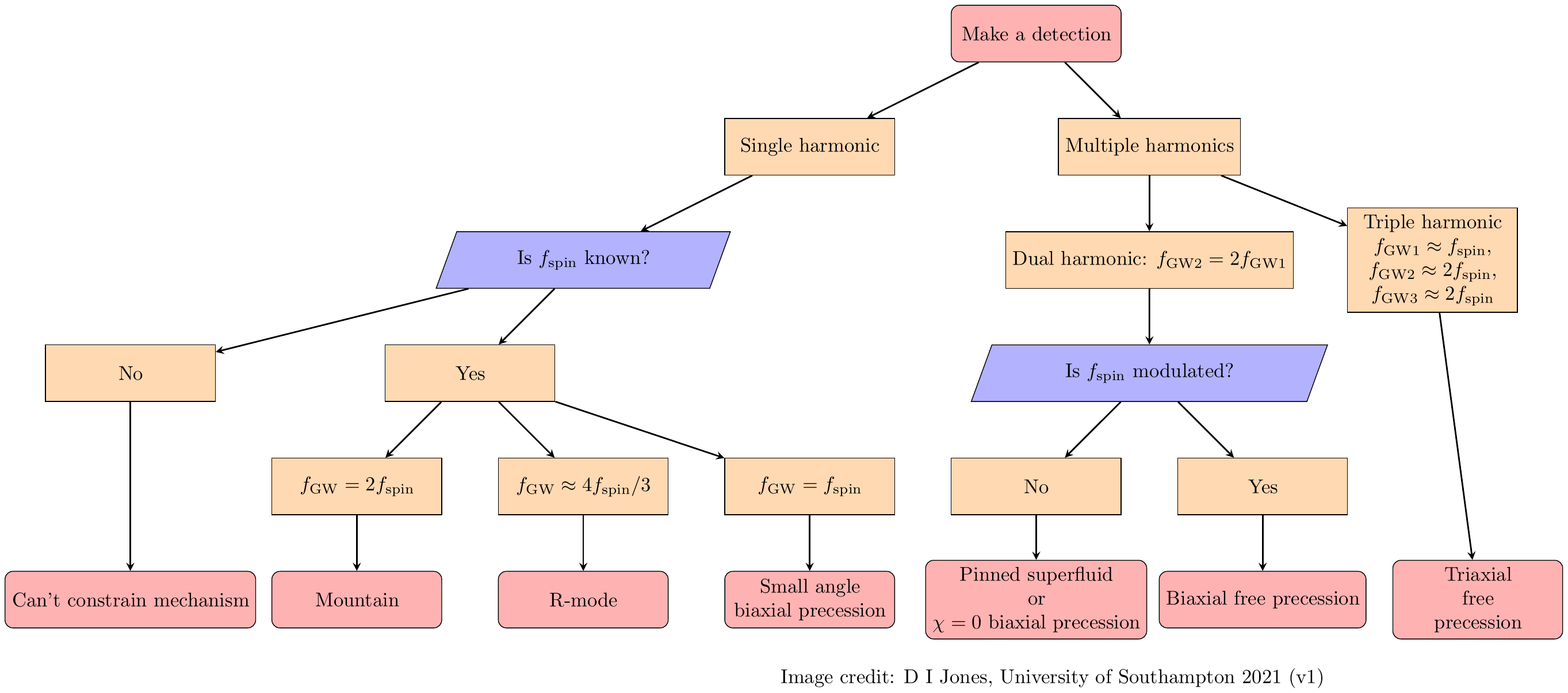}
\end{center}
\caption{Flow chart illustrating possible emission mechanisms, and how they relate the the frequency content of a CW detection.}
\label{fig:flow_chart}
\end{sidewaysfigure}

I have produced a flow chart to accompany this chapter; see Figure \ref{fig:flow_chart}.  The reader is invited to follow this flow chart when reading this article, to help see the relations between the various scenarios.

\section{Continuous wave emission mechanisms} \label{sect:mechanisms}

The simplest possible CW emission mechanism is that of a steadily and rigidly rotating neutron star with a non-axisymmetric mass distribution, i.e.\ a mountain \citep{st_83}.  For a star with angular velocity $\Omega_{\rm spin} = 2 \pi f_{\rm spin}$, at distance $r$ from Earth, with moment of inertia $I_3$ about the spin axis, and moments of inertia $I_1 \neq I_2$ about the two orthogonal axes, the gravitational wave amplitude can be parameterised by
\begin{equation}
\label{eq:h0_mountain}
h_0 = \frac{16\pi^2 G}{c^4} \frac{I_3 \epsilon_{21} f_{\rm spin}^2 }{r} , 
\end{equation}
where the asymmetry in the moment of inertia is  parameterised by
\begin{equation}
\label{eq:epsilon_def}
\epsilon_{21} = \frac{I_2-I_1}{I_3} ,
\end{equation}
a dimensionless quantity often known as the \emph{ellipticity}.  

A rotating  perfect fluid star in equilibrium would have $\epsilon_{21} = 0$.  Non-zero ellipticities require some sort of deforming force, either from elastic strain in the solid crust (and/or solid core, if such a component exists), or from magnetic strains \citep{gg_18_review}.  In either case, the gravitational wave emission is at twice the spin frequency, i.e. $f_{\rm GW}  = 2 f_{\rm spin}$.  In the case of elastic strains, the ellipticity is proportional to the strain $u$ in the crust, and is given approximately by \citep{ucb_00}:
\begin{equation}
\epsilon_{21} \sim 10^{-7} \frac{u}{10^{-2}} .
\end{equation}

In the case of magnetic deformations, the situation depends upon whether or not the interior proton fluid is in the  \emph{normal} or the \emph{superconducting state}.  If normal, the ellipticity is of order the magnetostatic energy divided by the gravitational binding energy:
\begin{equation}
\label{eq:epsilon_B_normal}
\epsilon_{21} \sim 10^{-12} \left(\frac{B_{\rm int}}{10^{12} \, \rm G}\right)^2 ,
\end{equation} 
where $B_{\rm int}$ is the \emph{internal} magnetic field strength;  see e.g. \citet{jone_02, lj_09}.  However, a few hundred years after birth, a neutron star will have cooled sufficiently for its protons to go into the superconducting state, in which case the ellipticity instead scales linearly in the internal magnetic field strength:
\begin{equation}
\label{eq:epsilon_B_supercon}
\epsilon_{21} \sim 10^{-8} \frac{B_{\rm int}}{10^{12} \, \rm G} ;
\end{equation}
see  \citep{land_12, land_14}.

The next most commonly considered emission mechanism is from fluid oscillations, specifically \emph{r-mode oscillations} \citep{pbr_81}.  The r-modes are a type of oscillation that only exist in rotating stars, with the restoring force being the Coriolis force.  They are favoured over other modes as they are considered to be the most likely mode to undergo the Chandrasekhar-Friedman-Schutz instability, where gravitational radiation reaction acts to amplify the mode \citep{ande_98, lom_98}.  The mode can be parameterised by a dimensionless amplitude $\alpha$, so that the gravitational waves can be parameterised by \citep{oetal_98, owen_10}:
\begin{equation}
\label{eq:h0_r-mode}
h_0 = \sqrt{\frac{8\pi}{5}} \frac{G}{c^5} \frac{1}{r} \alpha \omega^3 M R^3 \tilde J
\end{equation}
where $\omega$ is the gravitational wave (angular) frequency, $M$ and $R$ the mass and radius of the star, and $\tilde J$ a dimensionless constant hat depends upon stellar structure (see \citet{oetal_98} for details).

In the simplest approximation, where the star is modelled as a slowly rotating perfect fluid in Newtonian gravity, the oscillation frequency  of the relevant r-mode is simply $4/3$ times the spin frequency \citep{pbr_81}.   This is also the gravitational wave frequency, so we have $f_{\rm GW} = 4 f_{\rm spin} / 3$.  In more realistic treatments, other effects, principally rotation, elasticity and general relativity, combine to give a gravitational wave frequency slightly less than this \citep{lfa_03, ioj_15, cetal_19}.

These are the two most commonly considered emission mechanisms.  But there are others.  If a steadily spinning star is perturbed, it can be set into \emph{free precession}.  In this case, even a biaxial ($I_1 = I_2 \neq I_3$) star will radiate.  As first computed by \citet{zs_79}, there is then emission at two harmonically-related frequencies, which we can write as $\dot\phi$ and $2\dot\phi$:
\begin{align}
\label{eq:h0_phi}
h(\dot\phi) & \sim \frac{2 \dot\phi^2}{r} I_3 \epsilon_{31} \sin\theta \cos\theta , \\
\label{eq:h0_2phi}
h(2\dot\phi) & \sim \frac{2 \dot\phi^2}{r} I_3 \epsilon_{31} \sin^2\theta ,
\end{align}
where it is now a different sort of ellipticity that sources the emission:
\begin{equation}
\epsilon_{31} \equiv \frac{I_3 - I_1}{I_3} .
\end{equation}
Here, $\dot\phi$ is the angular velocity at which the star's $3$-axis moves in a cone around the invariant angular momentum axis $\bf J$, and $\theta$ is the opening angle of this cone, sometimes known as the \emph{wobble angle}; see \citet{ja_01} for a discussion.   We have stopped short of writing down proper equalities as we have not included various inclination angle factors that determine the polarisation content, as they are not important for our purposes (see \citet{zs_79} for the full equalities).

In this case, if the star is also visible as a radio pulsar, the relation between the frequency of the radio pulsations and the gravitational wave frequencies $(\dot\phi, 2\dot\phi)$ is slightly complicated, as described in \citet{ja_01, ja_02}.  The relation depends upon $\chi$, the angle between the star's $3$-axis and the ``magnetic'' axis $\bf m$  along which the pulsar beam is emitted.  If $\chi = 0$, the pulsar beam simply follows the $3$-axis, and the radio pulsations are emitted at the constant $\dot\phi$ frequency.  However, if $\chi \neq 0$, the situation is more complicated.  This is because the free precession of a biaxial body consists not just of the motion of the $3$-axis about $\bf J$; there is a slow superimposed rotation of the body about the (instantaneous position of the) $3$-axis at a rate $\dot\psi$, known as the ``body frame precession frequency'' or simply the precession frequency.  This precession frequency will be small in the realistic case of a nearly spherical ($|\epsilon_{31}| \ll 1$)  star:
\begin{equation}
\label{eq:psi_dot}
\dot\psi = - \epsilon_{31} \dot\psi \cos\theta .
\end{equation}
This produces a rather complex motion of the magnetic axis $\bf m$, as its position is determined by the sum of these two rotations: the fast rotation of the $3$-axis about $\bf J$ at angular velocity $\dot\phi$, and the slow superimposed rotation of $\bf m$ about the $3$-axis at the rate $\dot\psi$.  The observed frequency of the radio pulsations is then modulated once per free precession period $P_{\rm fp} = 2\pi/\dot\psi$, as is the observer's latitudinal cut through the pulsar beam.  Importantly for our purposes, the time-averaged radio pulsation frequency $\langle \nu_{\rm radio} \rangle$ will in general differ from the (constant) gravitational wave frequency $\dot\phi$, according to \citep{ja_01}:
\begin{align}
\label{eq:nu_radio_1}
2\pi \langle \nu_{\rm radio} \rangle &= \dot\phi + \dot\psi  \hspace{5mm} {\rm for} \hspace{5mm} \theta < \chi , \\
\label{eq:nu_radio_2}
2\pi \langle \nu_{\rm radio} \rangle &= \dot\phi  \hspace{11.5mm} {\rm for} \hspace{5mm} \theta > \chi .
\end{align}
This will play a role in the discussion below.

All of this applies to the free precession of a biaxial star.  In the more general case of a triaxial star $I_1 \neq I_2 \neq I_3$ the situation is more complex still, with both the radio pulsar and the gravitational wave frequencies being functions of time \citep{zimm_80}.  In the case of the small angle precession of a nearly spherical star, there is gravitational wave emission at three frequencies, which can be loosely identified with the $\dot\phi$, $2\dot\phi$ and $2\pi\langle \nu_{\rm radio} \rangle$ frequencies above \citep{zimm_80, vdb_05}.

As described above, the free precession of a star leads to the emission of more than one gravitational wave harmonic.  Also, in general, it produces modulation of the radio pulsar frequency.  The lack of observations of modulation in radio pulsar emission suggests that free precession is, at best, a rare phenomenon, which is why relative few gravitational wave searches have looked for such dual harmonic signals.  

There are two special cases where there is no modulation in the radio emission, but there is nevertheless dual harmonic gravitational wave emission.  The first was mentioned above: if the pulsar beam of a biaxial precessing star points exactly along the symmetry axis of the body, and if additionally the beam itself is symmetric about this axis, there will be no observed modulation.  This is clearly quite a special case.  

There is however one other scenario, where a steadily rotating star, with constant radio pulsation frequency,  can produce dual harmonic gravitational waves.  It is thought that all but the very youngest of neutron stars will be cool enough for some component of their neutron to be in a superfluid state \citep{ch_08}.  Such a superfluid rotates by forming an array of vortices, which can ``pin'' to the solid crust, preventing the associate angular momentum vector from moving relative to the crust.  The effect is as if a spinning gyroscope has been sewn into the crust \citep{shah_77}.  If this vortex array is misaligned with the principle axes of the crust's moment of inertia tensor, the non-precessing state of the star is such that, even though the rotation is steady, gravitational waves are produced at two harmonics, $f_{\rm spin}$ and $2f_{\rm spin}$ \citep{jone_10}.

In recognition of these two emission modes, a number of gravitational wave searches have been performed looking for precisely this dual harmonic emission, even in known pulsars with no sign of precession in their radio emission; see \citet{LVK_19_known_two_harmonics} for a recent example.

Another simpler but more speculative scenario can apply if the star has a solid core.  Such a core might consist of a solid quark phase, and have a very high shear modulus \citep{owen_10, gjs_12}.  If it is this core that is deformed, rather than the elastic crust, it will be the rotation rate of the core rather than the crust that fixed the gravitational wave frequency.  In principle, the two may rotate at different rates, thus opening up a new degree of freedom in the frequency spectrum \citep{LVK_08_beating_Crab}.

In recognition of the fact that both free precession and independently spinning deformed cores might induce a small off-set between the observed pulsar spin frequency and (a component of) the gravitational weave frequency, a number of so-called \emph{narrow band} gravitational wave searches have been performed; see \citet{LVK_19_known_narrowband} for a recent example.

\section{Detection of a single GW harmonic} \label{sect:single}

A CW detection might reveal the presence of just one gravitational wave component (``single harmonic''), or it may consist of multiple frequency components (``multi harmonic'').  Let us consider the single harmonic case first.

\subsection{Single GW harmonic: $f_{\rm spin}$ unknown} \label{sect:single_f_unknown}

Suppose a CW detection consists of just a single frequency component, and there is no associated electromagnetic counterpart, so we have no knowledge of $f_{\rm spin}$.  This is the worst-case scenario, in terms of learning about the emission mechanism.  We may have a mountain, or an r-mode.   In the mountain case, there is no way of inferring if it is of elastic or magnetic nature. 

There is one possible avenue for progress here, if the evolution in gravitational wave frequency of the spinning down star is fast enough for both the first and second time derivatives of $f_{\rm GW}$ to be measured.  In this case, assuming that the gravitational wave frequency is proportional to the spin frequency, one can take the definition of the \emph{braking index}, written in terms of the spin frequency:
\begin{equation}
 \label{eq:n}
 n \equiv \frac{f_{\rm spin} \ddot f_{\rm spin}}{\dot f_{\rm spin}^2} ,
 \end{equation}
 and evaluate it making use of the observed values of $f_{\rm GW}, \dot f_{\rm GW}, \ddot f_{\rm GW}$ in place of 
$f_{\rm spin}, \dot f_{\rm spin}, \ddot f_{\rm spin}$.  
 
The key consideration is this.  For a star spinning down entirely due to gravitational wave emission from a mountain of constant ellipticity $\epsilon_{21}$, we have $n=5$.  Similarly, for a star spinning down entirely due to gravitational wave emission from an r-mode of constant amplitude $\alpha$, we have $n=7$ \citep{gg_18_review}.  It follows that a measured value of $n$ close to one or other of these two values is strongly suggestive of (but does not prove) emission from the corresponding mechanism.
 
 Furthermore, in such a case, \citet{sj_21} have shown in detail that one can go some way to breaking the three-way degeneracy between the quantities $(I_3, r, \epsilon_{21})$ of equation (\ref{eq:h0_mountain}), or between the quantities $(I_3, r, \alpha)$ of equation (\ref{eq:h0_r-mode}), to infer combinations of only two of these quantities.  Given that $I_3$ is known to within a factor of $2$ or so from the modelling on neutron star structure for realistic equations of state (see discussion in \citet{wetal_18}, this allows estimation of $\epsilon_{21}$ or $\alpha$ to within a factor of $\sqrt{2}$ or so.  The distance $d$ can be similarly constrained, hence the `standard siren' analysis of \citet{sj_21}.  In this way, an estimate of $\epsilon_{21}$ or $\alpha$ can be obtained.  However, in the mountain case, there is still  no way of distinguishing between elastically and magnetically supported deformation.

There is one further possibility.  As described above, in general biaxial free precession leads to emission at a pair of harmonically-related frequencies, which we denoted as $\dot\phi$ and $2\dot\phi$; see equations (\ref{eq:h0_phi}) and  (\ref{eq:h0_2phi}).  However, in the (possibly most plausible) case of small wobble angle free precession, ie. $\theta \ll 1$, the amplitudes of the two components scale as $\theta$ and $\theta^2$, respectively; see equations (\ref{eq:h0_phi}) and  (\ref{eq:h0_2phi}), i.e. the emission at the lower harmonic dominates.  In such a case only a single component ($h_0(\dot\phi)$) might be detected.  So, we need to add small angle free precession of a biaxial star to our list of possibilities.

\subsection{Single GW harmonic: $f_{\rm spin}$ known}

In the event of a single harmonic CW detection where $f_{\rm spin}$ is also known, one can clearly use the relative values of $f_{\rm GW}$ and $f_{\rm spin}$ to make progress.  We will consider the most obvious possibilities.

\subsubsection{Single GW harmonic: $f_{\rm spin}$ known: $f_{\rm GW} \approx 2 f_{\rm spin}$}

In the case $f_{\rm GW} \approx 2 f_{\rm spin}$ we clearly have emission from a mountain.  If $f_{\rm GW} = 2 f_{\rm spin}$
exactly, or at least to within the accuracy of the gravitational wave measurement,  so would infer that the gravitational wave-producing deformation is tightly locked to the radio pulsation mechanism.  This is in fact what seems most likely from a theoretical standpoint.  The radio pulsations are almost certainly produced by particle acceleration in the (external) magnetic field; see e.g. \citet{lgs_12}.  The magnetic field itself is likely to remain tightly locked to the neutron star, as neutron star matter is thought to be of sufficiently high electrical conductivity that diffusive processes would act on timescales long compared with gravitational wave observations (see e.g.\ \citet{gr_92}).

Given this expectation, a small offset between $f_{\rm GW}$ and $f_{\rm spin}$ would be of great interest.  We can parameterise this offset in terms of a dimensionless quantity $\delta$:
\begin{equation}
\label{eq:delta_def}
f_{\rm GW} = 2 f_{\rm spin} (1 + \delta) .
\end{equation}
where $f_{\rm GW}/2$ is to be identified with the spin of the solid core, and $f_{\rm spin}$ with the spin of the crust.

A small positive value of $\delta$ could be interpreted as gravitational wave emission from a spinning down sold core, that is imperfectly coupled to the rest of the star \citep{LVK_08_beating_Crab}.  The picture would be that the external electromagnetic spin-down torque acts directly on the crust, and whatever is coupled tightly to the crust, but not directly on the solid core.  We will denote the moments of inertia of these two components as $I_{\rm crust}$ and $I_{\rm core}$, so that the total is  $I = I_{\rm crust} + I_{\rm core}$.    The coupling torque between the crust and core may be of frictional nature, acting so as to restore co-rotation between the two on timescales short compared with the spin-down timescale, with a magnitude proportional to the spin frequency difference.  We could then write the coupling torque acting on the crust as 
\begin{equation}
T_{\rm coupling} = -A (f_{\rm crust} - f_{\rm core}) ,
\end{equation}
where $A$ is some positive constant that parameterises the strength of the coupling torque.  There will be an equal but opposite coupling torque $-T_{\rm coupling}$ acting on the core.  In such a situation, if the spin-down torque due to the gravitational wave emission were to be neglected, one can easily show that the fractional frequency off-set between the crust and core is given by the ratio of the crust-core coupling timescale to the spin-down timescale:
\begin{equation}
\label{eq:delta_coupling}
\frac{f_{\rm GW}/2 - f_{\rm spin}}{f_{\rm spin}} = \delta \approx \frac{\tau_{\rm coupling}}{\tau_{\rm spin-down}} \frac{I}{I_{\rm crust}}.
\end{equation}
In this way, one could infer not only the existence of the sold core, but also the strength of its coupling with the rest of the star.

A variation on this scenario would be that the angular momentum exchange between the spinning down crust and solid core is not perfectly smooth, but rather has a random component to it, that might be related to the seemingly ubiquitous phenomenon of pulsar \emph{timing noise} \citep{lgs_12}.  In this case, small upward fluctuations in $f_{\rm spin}$ would, by angular momentum conservation, be accompanied by small downward fluctuations in $f_{\rm GW}$ (and \emph{vice versa}), with a constant of proportionality related to the relative moments of inertia of the two components \citep{jone_04}.  Identification of such a correlation by comparison of the electromagnetic and gravitational wave data streams would provide a novel insight into the interior physics of neutron stars.

Alternatively, a small but non-zero value of $\delta$ could be interpreted as due to free precession \citep{ja_02, LVK_08_beating_Crab}.  However, in this case, emission at other gravitational harmonics is to be expected, and the radio pulsar frequency should be modulated in time.  The absence of such features would favour the solid core interpretation.

Regardless of these issues, it is of interest to ask if there is any way of discriminating between elastically and magnetically supported mountains.  This is a tricky degeneracy to break.  One possibility applies if the distance to the source is known, say from electromagnetic means (e.g.\ radio pulsar dispersion measure; \citep{lgs_12}).  In that case the value of the ellipticity $\epsilon_{21}$ can be estimated (again, only up the uncertainty in $I_3$, about a factor of $2$), and the total contribution of gravitational wave emission  to the total spin-down rate inferred.  If this is small, so that most of the spin-down is from electromagnetic emissions, one can use the standard magnetic dipole formula to estimate the strength of the \emph{external} magnetic field $B_{\rm ext}$ \citep{st_83}.   This could be compared with the strength that would be inferred from the \emph{internal} magnetic field $B_{\rm int}$, required to support the mountain, if of magnetic origin, as per equations (\ref{eq:epsilon_B_normal}) and (\ref{eq:epsilon_B_supercon}) above.    One could then compute the ratio
\begin{equation}
\label{eq:r_mag_def}
r_{\rm mag} \equiv \frac{B_{\rm ext}}{B_{\rm int}} .
\end{equation}
The actual value of this ratio in neutron stars is not known, but modelling suggests that the internal field is likely to be somewhat stronger than the external one, see e.g. \citet{lj_09}.  It follows that a value of $r_{\rm mag}$ greater than unity, or a value much less than unity, would make the magnetic mountain interpretation less plausible, favouring the elastic strain interpretation.

\subsubsection{Single GW harmonic: $f_{\rm spin}$ known: $f_{\rm GW} \approx 4 f_{\rm spin}/3$}

In the case of a single gravitational wave component with $f_{\rm GW} \approx 4 f_{\rm spin}/3$ we clearly have r-mode emission.  In this case the departure of $f_{\rm GW}$ from $4f_{\rm spin}/3$ encodes useful information, as is discussed in some detail in \citet{lfa_03, ioj_15, cetal_19}.  The largest contribution to this frequency shift is likely to be due to the effects of general relativity, and to a good approximation allows one to infer the star's compactness $M/R$.
However, care must be taken when making this argument.  There are likely to be narrow bands in $f_{\rm spin}$ where there is a near resonance between the r-mode frequency and the frequencies of other normal modes, and this can have an even larger effect of the r-mode (and hence gravitational wave) frequency than the relativistic effects.  This has been shown explicitly for the resonance of the r-mode with torsional oscillations of the elastic crust \citep{lu_01}.  Resonances with superfluid modes can also play a role, and may be important in determining the spin frequencies of some accreting neutron stars \citep{kgc_16}

\subsubsection{Single GW harmonic: $f_{\rm spin}$ known: $f_{\rm GW} \approx f_{\rm spin}$}

In the case $f_{\rm GW} \approx f_{\rm spin}$ neither mountains nor r-modes can be at work.  As described in Section \ref{sect:single_f_unknown}, the natural interpretation would be small-angle free precession of a biaxial body, i.e. the $\theta \ll 1$ case of equations (\ref{eq:h0_phi}) and  (\ref{eq:h0_2phi}).  Given that $\theta \ll 1$, we would presumably be in the regime where $\theta < \chi$, where $\chi$ is the angle between the biaxial body's symmetry axis and the magnetic dipole axis $\bf m$.  In this case, equation (\ref{eq:nu_radio_1}) applies, and the fractional difference between $f_{\rm GW}$ ($= \dot\phi / (2\pi)$ in the notation of Section \ref{sect:mechanisms}),  and $f_{\rm spin}$ can be used to infer the relevant ellipticity $\epsilon_{31}$, as follows from combining equations (\ref{eq:nu_radio_1}) and (\ref{eq:psi_dot} to give:
\begin{equation}
\epsilon_{31} \approx \frac{f_{\rm GW} - f_{\rm spin}}{f_{\rm GW}} ,
\end{equation}
where we have used the small $\theta$, small $\epsilon_{31}$ approximations.

\section{Detection of  multiple GW harmonics} \label{sect:multi}

In the case of emission at multiple gravitational wave harmonics, it is easier to make progress, even in the absence of an observed pulsar spin frequency.  We will label the frequencies of the gravitational wave components as $f_{\rm {GW}1}, f_{\rm {GW}2}, \dots$.

\subsection{Dual GW harmonic: $f_{{\rm GW}1}$, $f_{{\rm GW}2} = 2f_{\rm {GW}1}$}

In the case of dual harmonic emission, with the upper harmonic at twice the frequency of the lower ($f_{{\rm GW}2} = 2f_{\rm {GW}1}$), biaxial free precession is indicated, as described by equations (\ref{eq:h0_phi}) and  (\ref{eq:h0_2phi}) \citep{zs_79, ja_02}.  The relative amplitude of the harmonics contains information on the wobble angle $\theta$.    In the absence of a measurement of $f_{\rm spin}$, one cannot go further.  However, if $f_{\rm spin}$ is measured, then one can draw different conclusions depending upon whether it is modulated in time or not.

\subsubsection{Dual GW harmonic: $f_{{\rm GW}1} = f_{\rm spin}$, $f_{{\rm GW}2} = 2f_{\rm spin}$, with $f_{\rm spin}$ not modulated}

If the radio pulsation is measured to be steady, and equal to the lower of the two (harmonically-related) gravitational wave frequencies,  then there are two possibilities.  The first is that biaxial free precession is taking place, with $\chi=0$, i.e.\ the radio emission is taking place along the symmetry axis of the biaxial body, and so the magnetic axis is simply carried around at the frequency $\dot\phi$ that appears in equations (\ref{eq:h0_phi}) and  (\ref{eq:h0_2phi}) \citep{ja_01}.  The second possibility is that superfluid pinning is at work, misaligned with the principle axes of the star's moment of inertia tensor, giving a perfectly steady $f_{\rm spin}$, but nevertheless emitting gravitational waves at both $f_{\rm spin}$ and $2f_{\rm spin}$ \citep{jone_10}.

\subsubsection{Dual GW harmonic: $f_{{\rm GW}1} = f_{\rm spin}$, $f_{{\rm GW}2} = 2f_{\rm spin}$, with $f_{\rm spin}$ modulated}

in the case of emission at two harmonically related frequencies, if the radio pulsation frequency is modulated, free precession of a biaxial body with $\chi \neq 0$ is indicated.  The time-averaged radio frequency is then given by equation (\ref{eq:nu_radio_1}) or (\ref{eq:nu_radio_2})  above, depending upon the relative values of $\theta$ and $\chi$.  Either way, the observed periodicity of the modulations in $f_{\rm spin}$ gives the free precession frequency $P_{\rm fp}$, which in turn can be used to estimate the degree of biaxiality $\epsilon_{31}$ via equation (\ref{eq:psi_dot}).

\subsection{Triple GW harmonic: $f_{{\rm GW}1} \approx f_{\rm spin}$, $f_{{\rm GW}2} \approx 2f_{\rm spin}$, $f_{{\rm GW}3} \approx 2f_{\rm spin}$}

In the case of gravitational wave emission from a triplet of frequencies, of the approximate form $(f_{\rm spin}, 2f_{\rm spin}, 2f_{\rm spin})$, small angle ($\theta \ll 1$) free precession of a triaxial star is indicated, for a star with a small degree of biaxiality ($|\epsilon_{31}| \ll 1$) and and even smaller degree of triaxiality ($|\epsilon_{21}| \ll |\epsilon_{21}|$); see \citet{zimm_80} and \citet{vdb_05}.  In this case, the emission is close to a sum of what one would expect from biaxial precession, plus a mountain, with $f_{{\rm GW}1}$ and (say) $f_{{\rm GW}2}$ sourced mainly by $\epsilon_{31}$, and $f_{{\rm GW}3}$ sourced by $\epsilon_{21}$.  In this case, the relative amplitudes of the three harmonics, and their separation in frequency, would give information on the degree of triaxiality; see \citet{zimm_80} and \citet{vdb_05} for discussion.

\section{Summary and reflections} \label{sect:summary}

Clearly, there are many different scenarios when it comes to the frequency content of a CW signal.  I have sketched out here what I believe to be some of the more plausible possibilities, and attempted to indicate how we can use this information to draw conclusions about some parameters of the star.   I hope that in the event of an eventual, and, let us hope not far off, first detection, one or other of these many possibilities can be identified, allowing us to extract as much information about the physics of the neutron star as possible.

\section*{Acknowledgements}

I would like to acknowledge many useful conversations with colleagues in the Southampton Gravity Group, the LIGO Scientific Collaboration, and the Virgo Collaboration, on many of the ideas discussed here, including comments on this chapter from Bryn Haskell.  I also acknowledge financial support from the Science and Technology Facilities Council (STFC, UK) via grant no. ST/R00045X/1.


\end{document}